\documentclass{mn2e}
\usepackage{natbib}
\usepackage{graphicx}
\usepackage{pslatex}

\begin{document}

\title[Depolarisation through the cosmic ages]{The depolarisation properties of powerful extragalactic radio sources as a function of cosmic epoch}

\author[Goodlet \& Kaiser]{J.A.Goodlet\thanks{email: j.goodlet@talk21.com} \& C. R. Kaiser\\ 
School of Physics \& Astronomy, University of Southampton, Southampton SO17 1BJ
}

\maketitle

\begin{abstract}
We use the observed polarisation properties of a sample of 26 powerful radio galaxies and radio-loud quasars to constrain the conditions in the Faraday screens local to the sources. We adopt the cosmological redshift, low-frequency radio luminosity and physical size of the large-scale radio structures as our `fundamental' parameters. We find no correlation of the radio spectral index with any of the fundamental parameters. The observed rotation measure is also independent of these parameters suggesting that most of the Faraday rotation occurs in the Galactic foreground. The difference between the rotation measures of the two lobes of an individual source as well as the dispersion of the rotation measure show significant correlations with the source redshift, but not with the radio luminosity or source size. This is evidence that the small-scale structure observed in the rotation measure is caused by a Faraday screen local to the sources. The observed asymmetries between the lobes of our sources show no significant trends with each other or other source properties. Finally, we show that the commonly used model for the depolarisation of synchrotron radio emission by foreground Faraday screens is inconsistent with our observations. We apply alternative models to our data and show that they require a strong increase of the dispersion of the rotation measure inside the Faraday screens with cosmological redshift. Correcting our observations with these models for redshift effects, we find a strong correlation of the depolarisation measure with redshift and a significantly weaker correlation with radio luminosity.  We did not find any (anti-)correlation of depolarisation measure with source size. All our results are consistent with a decrease in the order of the magnetic field structure of the Faraday screen local to the sources for increasing cosmological redshift.
\end{abstract}


\section{Introduction}

The large-scale radio structure of radio galaxies and radio-loud quasars is created by the interaction of powerful jets with the gaseous environment of their host galaxies. The jets themselves are produced in the immediate vicinity of supermassive black holes at the centres of the host galaxies. Because radio radiation is not easily obscured, luminous radio-loud AGN are in principle ideal objects to study the co-evolution of supermassive black holes and their large-scale environments throughout the universe. However, determining the properties of the gas in the surroundings of these objects is not straightforward and a number of techniques have been applied in the past. 

Galaxy counts around individual host galaxies can be used to determine the galactic environment, field, group or cluster, of the radio source \citep[e.g.][]{hl91,wll00}. Obviously the density of galaxies does not directly yield a constraint on the properties of the gas in between the galaxies. A relation of galaxy to gas density must be assumed. 

The properties of the large-scale radio structure are shaped by the density of the gas the jets are ploughing into. The redshift distribution of, for example, the linear sizes of the lobes of radio galaxies of type FRII \citep{fr74} in complete samples can be used to constrain the properties of the source environments \citep[e.g.][]{ka98a,brw99}. Detailed fitting of two-dimensional radio maps can also be applied to individual sources \citep{ck00b}. However, the determination of the gas density from the source size is always model-dependent. Although most models for the evolution of radio lobes employ some form or other of ram pressure equilibrium between the jets and the receding gas in the source environment, the model details introduce considerable uncertainty in the estimates of the gas density.

The gas surrounding the lobes of radio galaxies consists of a magnetised plasma which acts as a Faraday screen for the radio synchrotron emission of the lobes. By studying the polarisation properties of the radio emission, we can investigate the properties of this screen. The observable rotation measure caused by the Faraday screen depends only on the density of the plasma and also on the strength and orientation of the magnetic field in this material. Despite this degeneracy, we can hope to probe the properties of the gas in the source environments directly. The aim of the present paper is to use this tool to study a sample of 26 powerful radio sources defined in \citet[][hereafter G04]{gkb03}. In order to break the degeneracy due to Malmquist bias between radio-luminosity and redshift observed in previous studies \citep[e.g.][]{kcg72,lpr86}, we defined 3 subsamples
of sources chosen from the 3C and 6C/7C catalogues. This selection enables us to investigate any trends of the polarisation properties of our sources with redshift and/or radio luminosity independently.

In Section \ref{stats} we briefly describe our sample and the statistical methods we use. We also present the observational data that forms the basis for the subsequent analysis. Section \ref{fundamentals} investigates trends of the polarisation properties with `fundamental' source parameters such as redshift, radio luminosity and linear size. Asymmetries between the properties of the two lobes of each source are discussed in Section \ref{asym}. In Section \ref{tribble} we show that the polarisation properties of our sources are inconsistent with the commonly used model by \citet{bb66} for external Faraday screens. We then apply the more sophisticated models proposed by \citet{pt91} to our data. Finally, Section \ref{conc} summarises our main conclusions.

Throughout this paper we use a cosmological model with $H_0=75$\,km\,s$^{-1}$\,Mpc$^{-1}$, $\Omega _{\rm m} =0.35$ and $\Omega _{\Lambda} =0.65$.

\section{Data and Statistical methods}
\label{stats}

\subsection{Sample description and observables}

Our sample of 26 sources is divided into three subsamples. Sample A was
defined as a subsample chosen from the 6CE \citep{erl97} subregion of
the 6C survey \citep{hmw90}, and the 7C III subsample
\citep{lrh99}, drawn from the 7C survey \citep{pwr98}. The
selected sources have redshifts $0.8<z<1.3$, and radio luminosities at 151\,MHz of $6.5 \times 10^{26} {\rm \,W\,Hz^{-1}}<P_{151 {\rm MHz}}<1.35
\times 10^{27} {\rm \,W\,Hz^{-1}}$. Sample B was defined as a subsample
from the revised 3CRR survey by \citet{lrl83} containing sources
within the same redshift range but with luminosities in the range $6.5
\times 10^{27} {\rm \,W\,Hz^{-1}}<P_{151 {\rm MHz}}<1.35 \times 10^{28}
{\rm \,W\,Hz^{-1}}$. Sample C is also from the 3CRR catalogue; it has the
same radio luminosity distribution as sample A, but with $0.3<z<0.5$. For full details on the sample selection and the
data reduction see G04. 

In the following we will
use source properties derived from the observations: Spectral index,
rotation measure and depolarisation. All these properties were
averaged over individual lobes.  We define the spectral index,
$\alpha$, by $S_{\nu}
\propto
\nu^{\alpha}$. The depolarisation measure, $DM$, is defined as the ratio of the
percentage polarised flux at 4.8\,GHz and the percentage polarised flux
at 1.4\,GHz. The rotation measure, $RM$, is related to the
degree of rotation of the polarisation position angle over a set
frequency range, in our case from 1.4\,GHz to 4.8\,GHz. All source properties used here were determined for individual sources in G04.

The angular resolution of our observations is limited by the need for information on the position angle of the plane of polarisation at 1.4\,GHz. For sources with lobes where the degree of polarisation is uniform and the polarisation angle does not change on small scales, we can be confident that our measured depolarisation measures are accurate. However, some sources may exhibit polarisation structure on small angular scales below our resolution limit and for these sources our measurements will be affected by beam depolarisation. Comparing our maps for the sources at low redshift presented in G04 with the higher resolution maps of \citet{grh04}, obtained mainly at 8\,GHz, we find a few sources where beam depolarisation may play a role. Examples include 3C\,351 and the northern lobe of 3C\,46. Our general impression for this subsample is that sources with distorted lobe structures may suffer from beam depolarisation. This should be borne in mind in the following analysis. Unfortunately, in the absence of better data, particularly for the sources at higher redshifts, it is not possible to quantify this effect.

The difference in the spectral index, depolarisation and rotation
measure over the lobes is given by d$\alpha$, d$DM$ and d$RM$
respectively. All differential properties are taken to be the
difference of the averages over individual lobes. The rms-variation or dispersion of
the rotation measure across a lobe is defined by $\sigma_{\rm RM}$. 

In G04
we argue that the average rotation measure of the sources, $RM$, is
mainly caused by the interstellar medium in the Galaxy rather than the
material in the vicinity of our sources. However, fluctuations in the
Faraday screen in the source environment will give rise to the
observed variations of $RM$ across the source on smaller spatial scales as measured by d$RM$ and $\sigma_{\rm RM}$, as well as the depolarisation, $DM$. Thus we
need to correct for the redshift dependence of these
quantities. We correct d$RM$ and $\sigma_{\rm RM}$ to the sources' frame
of reference by multiplying the measured values by $\left( 1 + z\right)^2$. Note that this factor only corrects for the wavelength dependence of $RM$. It does not correct for the resolution effects discussed in Section \ref{tribble}. We also adjust $DM$ and d$DM$ to their values appropriate for a
uniform cosmological redshift $z=1$. The latter adjustment requires a model for the structure of the Faraday screen local to the source which gives rise to these measurements. We initially apply the commonly used model presented in \citet[][hereafter B66]{bb66}, but discuss alternative, potentially more accurate models that also affect our measurements of $\sigma _{\rm RMz}$ in Section \ref{tribble}. The adjusted quantities are labelled d$RM$$_z$, $\sigma_{\rm RMz}$, $DM$$_z$ and d$DM$$_z$. We summarise the observed properties of all sources in our sample in Table \ref{sources}. 

\begin{table*}
\begin{tabular}{cccccccccc}
Name & $z$ & $P_{151}$ / W\,Hz$^{-1}$ & ID & $D$ / kpc & $S_{4.8}$ / mJy & $\alpha$ & $RM$ / rad\,m$^{-2}$ & $\sigma _{\rm RMz}$ / rad\,m$^{-2}$ & $DM_z$ \\
\hline
6C\,0943+39 & 1.04 & $1.0\times10^{27}$ & W & 56 & 31.0 & -0.72 & 1.6 & 67.0 & 2.36\\
 &  &  & E & 45 & 42.0 & -1.01 & -19.1 & 422.0 & 4.38 \\[1ex]
6C\,1011+36 & 1.04 & $1.1\times10^{27}$ & N & 175 & 44.3 & -0.88 & 30.8 & 53.3 & 1.60 \\
 &  &  & S & 220 & 18.7 & -0.89 & 12.6 & 42.0 & 1.07 \\[1ex]
6C\,1018+37 & 0.81 & $8.8\times10^{26}$ & NE & 279 & 46.4 & -0.85 & 0.85 & 20.0 & 1.15\\
 &  &  & SW & 294 & 28.7 & -0.84 & 11.2 & 9.5 & 1.82 \\[1ex]
6C\,1129+37 & 1.06 & $1.1\times10^{27}$ & NW & 63 & 46.5 & -0.85 & -19.3 & 229.2 & 2.86 \\
 &  &  & SE & 69 & 73.2 & -0.90 & 0.0 & 93.4 & 6.35 \\[1ex]
6C\,1256+36 & 1.07 & $1.3\times10^{27}$ & NE & 69 & 57.8 & -0.79 & 5.9 & 93.4 & 1.42\\
 &  &  & SW & 62 & 101.4 & -0.87 & 15.4 & 60.0 & 1.01 \\[1ex]
6C\,1257+36 & 1.00 & $1.1\times10^{27}$ & NW & 134 & 43.5 & -0.71 & -115.3 & 40.0 & 1.27 \\
 &  &  & SE & 150 & 20.5 & -1.06 & -115.6 & 64.0 & 1.90 \\[1ex]
7C\,1745+642 & 1.23 & $9.5\times10^{26}$ & N & 66 & 23.5 & -0.86 & --- & --- & 2.71 \\
 &  &  & S & 99 & 33.8 & -0.91 & 12.8 & 15.4 & 1.00 \\[1ex]
7C\,1801+690 & 1.27 & $1.0\times10^{27}$ & N & 89 & 8.8 & -0.97 & 44.8 & 82.5 & 1.19 \\
 &  &  & S & 113 & 28.7 & -0.81 & 20.8 & 51.5 & 1.73 \\[1ex]
7C\,1813+684 & 1.03 & $7.1\times10^{26}$ & NE & 282 & 15.0 & -0.92 & 13.9 & 350.3 & 1.02 \\
 &  &  & SW & 152 & 30.2 & -0.79 & -68.4 & 171.8 & 1.15 \\[1ex]
3C\,65 & 1.18 & $1.0\times10^{28}$ & W & 83 & 524.0 & -1.00 & -82.6 & 101.7 & 6.12 \\
 &  &  & E & 68 & 240.9 & -1.03 & -86.1 & 46.1 & 1.42 \\[1ex]
3C\,68.1 & 1.24 & $1.1\times10^{28}$ & N & 178 & 667.2 & -0.82 & -26.6 & 139.0 & 1.45 \\
 &  &  & S & 223 & 36.8 & -1.08 & 57.9 & 300.1 & 2.69 \\[1ex]
3C\,252 & 1.11 & $7.2\times10^{27}$ & NW & 159 & 178.7 & -1.00 & 15.7 & 89.9 & 1.10 \\
 &  &  & SE & 257 & 80.0 & -1.10 & 58.5 & 202.1 & 2.50 \\[1ex]
3C\,265 & 0.81 & $6.5\times10^{27}$ & NW & 331 & 224.0 & -0.62 & 42.2 & 71.8 & 1.52 \\
 &  &  & SE & 216 & 318.9 & -0.78 & 32.8 & 74.7 & 1.35 \\[1ex]
3C\,268.1 & 0.97 & $1.0\times10^{28}$ & E & 151 & 262.3 & -0.92 & 21.7 & 41.9 & 1.52 \\
 &  &  & W & 194 & 2296.6 & -0.58 & 26.8 & 51.6 & 1.82 \\[1ex]
3C\,267 & 1.14 & $1.0\times10^{28}$ & E & 158 & 184.0 & -1.17 & -9.6 & 111.7 & 1.43 \\
 &  &  & W & 129 & 479.2 & -0.83 & -21.5 & 107.2 & 1.08 \\[1ex]
3C\,280 & 1.00 & $1.2\times10^{28}$ & E & 51 & 326.0 & -1.01 & -37.7 & 80.4 & 1.83 \\
 &  &  & W & 61 & 1289.2 & -0.76 & -7.5 & 6.0 & 1.54 \\[1ex]
3C\,324 & 1.21 & $1.3\times10^{28}$ & NE & 50 & 432.6 & -1.05 & 22.1 & 112.3 & 2.15 \\
 &  &  & SW & 52 & 166.0 & -1.14 & 43.0 & 82.0 & 2.75 \\[1ex]
4C\,16.49 & 1.29 & $9.9\times10^{27}$ & N & 89 & 120.0 & -0.88 & -4.3 & 295.8 & 1.10 \\
 &  &  & S & 51 & 142.3 & -1.32 & 0.9 & 299.8 & 1.16 \\[1ex]
3C\,16 & 0.41 & $7.4\times10^{26}$ & SW & 200 & 484.9 & -0.97 & -4.3 & 34.6 & 1.16 \\
 &  &  & NE & 187 & 22.1 & -0.87 & --- & --- & 1.22 \\[1ex]
3C\,42 & 0.40 & $7.5\times10^{26}$ & NW & 87 & 353.9 & -0.87 & -2.4 & 27.6 & 1.01 \\
 &  &  & SE & 92 & 450.6 & -0.86 & 5.0 & 27.8 & 1.00 \\[1ex]
3C\,46 & 0.44 & $7.9\times10^{27}$ & NE & 512 & 162.7 & -0.94 & -4.8 & 8.1 & 1.32 \\
 &  &  & SW & 376 & 173.7 & -0.91 &-2.9  & 8.7 & 1.17 \\[1ex]
3C\,299 & 0.37 & $6.9\times10^{26}$ & NE & 71 & 876.5 & -0.93 & -126.3 & 142.5 & 1.15 \\
 &  &  & SW & 36 & 53.7 & -0.71 & 16.0 & 12.9 & 1.04 \\[1ex]
3C\,341 & 0.45 & $8.8\times10^{26}$ & NE & 280 & 123.8 & -0.85 & 20.3 & 21.0 & 1.11 \\
 &  &  & SW & 145 & 265.9 & -1.00 & 18.2 & 25.2 & 1.13 \\[1ex]
3C\,351 & 0.37 & $7.6\times10^{26}$ & NE & 158 & 1093.9 & -0.80 & 1.0 & 23.5 & 1.02 \\
 &  &  & SW & 188 & 77.1 & -0.93 & 4.4 & 20.3 & 1.24 \\[1ex]
3C\,457 & 0.43 & $9.6\times10^{27}$ & NE & 568 & 208.6 & -1.02 & --- & --- & 1.00 \\
 &  &  & SW & 519 & 290.5 & -0.94 & --- & --- & 1.05 \\[1ex]
4C\,14.27 & 0.39 & $8.8\times10^{26}$ & NW & 103 & 107.0 & -1.06 & -13.0 & 10.4 & 1.09 \\
 &  &  & SE & 76 & 124.8 & -1.17 & -17.3 & 9.3 & 1.01 \\
\hline
\end{tabular}
\caption{Observational properties of the sources in our sample. $P_{151}$ is the radio luminosity at 151\,MHz, $D$ the physical size of individual lobes and $S_{4.8}$ is the radio flux at 4.8\,GHz. $\sigma_{\rm RMz}$ is the dispersion of the rotation measure corrected for redshift effects. $DM_z$ is the depolarisation adjusted to a common redshift of $z=1$ using the model of \citet{bb66}. All data taken from G04.}
\label{sources}
\end{table*}

\subsection{Statistical methods}

In this paper we use the Spearman Rank test (SR), the Partial Spearman Rank test (PSR) and Principal Component Analysis (PCA). SR is a non-parametric correlation test which assigns a rank to given source properties, say X and Y, and computes the coefficient of the correlation, $r_{\rm XY}$, between the ranks. By definition $-1 \le r_{\rm XY} \le 1$, where negative values of $r_{\rm XY}$ indicate an anti-correlation. The null hypothesis that no correlation is present corresponds to $r_{\rm XY} =0$ and in general $r_{\rm XY}$ has a Student-t distribution. A (anti-)correlation with a confidence level exceeding 95\% will result in a Student-t value greater than 2 for the size of our sample \citep[e.g.][]{rhb97}.

PSR tests whether a correlation between properties X and Y is due to independent individual correlations of X and Y with a third property Z \citep[e.g.][]{jm82}. The correlation coefficient, $r_{\rm XY,Z}$, increases (decreases) from zero if the correlation (anti-correlation) is predominantly between X and Y, independent of Z. The positive significance of the (anti-)correlation is measured by $D_{\rm XY,Z}$. 

PCA is a multi-variate statistical test \citep[e.g.][]{td64,ef84}. PCA is a
linear self--orthogonal transformation from an original set of $n$
objects and $m$ attributes forming a $n \times m$ matrix, with zero mean
and unit variance, to a new set of parameters, known as principal
components. The principal components of the new dataset are all
independent of each other and hence orthogonal. The first component
describes the largest variation in the data and has, by definition,
the largest eigenvalue of the $n \times m$ matrix. In physical terms
the magnitude of the eigenvalue determines what fraction of the
variance in the data any correlation describes. Thus a strong
correlation will be present in the first eigenvector and will only
strongly reverse in the last. PCA in its simplest terms is an
eigenvector--eigenvalue problem on a transformed, diagonalised and
standardised set of variables. $n$ objects will create an
eigenvector--eigenvalue problem in $n-1$ dimensions. To avoid problems
with the interpretation of our results, we do not use more than 4
parameters in any part of the analysis.

\section{Trends with redshift, luminosity and size}
\label{fundamentals}

The aim of our analysis is to investigate any trends with redshift and/or radio luminosity of observables that may depend on the properties of the gas surrounding the radio lobes of our sample sources. Another important parameter that may influence our results is the physical size of the radio lobes. Because we expect significant gradients for the gas density in the typical environments of sources, the properties of small sources will clearly be shaped by very different conditions than those of large sources, even if the overall gaseous environment is the same. Many studies have investigated the dependence of the average size of sources as a function of redshift and radio luminosity in complete samples of sources \citep[e.g.][]{as88,bm88,vk89,as93,ner95,ka98a,brw99}. This is not the purpose of the study presented here. Our sample is not complete and sources were selected in order to cover an equal range of physical sizes within our sample limits of redshifts and radio luminosities. 

In the following we will refer to the redshift, $z$, radio luminosity at 151\,MHz, $P$, and the physical size, $D$, as the `fundamental' source parameters. In order to better understand the correlations of other observables with the fundamental parameters, we first need to investigate any correlations between them. This analysis also serves as a useful test of the robustness of our sample selection.

\begin{table}
\begin{center}
\begin{tabular}{cccccc}
X,Y & $r_{\rm XY}$ & Student-t & Z & $r_{\rm XY,Z}$ & $D_{\rm XY,Z}$\\
\hline
$z$, $P$ & 0.70 & 4.8 & $D$ & 0.69 & 4.0\\
$z$, $D$ & -0.23 & -1.2 & $P$ & 0.04 & 0.20\\
$P$, $D$ & -0.13 & -0.64 & $z$ & -0.20 & 0.93\\
\hline
\end{tabular}
\end{center}
\caption{Spearman Rank coefficients for correlations between the fundamental parameters.}
\label{fund}
\end{table}

Table \ref{fund} illustrates the relations between the fundamental parameters in our sample. Clearly, redshift and radio luminosity are strongly correlated, but this correlation simply follows from our sample selection. Using the statistical techniques described above, we can isolate this correlation from others. In fact, it provides us with a `bench mark' for other correlations. 

The SR results also imply a weak anti-correlation between the redshift of a source and its physical size. However, the PSR results show that the relation between the three fundamental parameters is more complex than a simple anti-correlation between $z$ and $D$. We also found that the anti-correlation disappears if the two exceptionally large sources 3C\,46 and 3C\,457, both at low redshifts, are removed from the sample. In practice, the $z$-$D$ anti-correlation is too weak to introduce any significant bias into our results. We therefore conclude that our sample contains a sufficient range of physical sizes at each redshift and radio luminosity.

In the following we investigate not only the observed source properties, but also differences of these properties between the two lobes of each source. Clearly there are twice as many direct measurements of, say, the spectral index of individual lobes, than measurements of the difference of the spectral index between the lobes. To avoid complications with the different number of measurements, we always use values for all source properties averaged over the entire source.

\subsection{Spectral index}

\begin{table}
\begin{center}
\begin{tabular}{cccccc}
X, Y & $r_{\rm XY}$ & Student-t & X, Y & $r_{\rm XY}$ & Student-t \\
\hline
$\alpha$, $z$ & -0.13 & -0.62 & d$\alpha$, $z$ & 0.15 & 0.77\\
$\alpha$, $P$ & 0.04 & 0.19 & d$\alpha$, $P$ & 0.27 & 1.35\\
$\alpha$, $D$ & 0.21 & 1.05 & d$\alpha$, $D$ & -0.21 & -1.05\\
\hline
\end{tabular}
\end{center}
\caption{Spearman Rank coefficients for the correlations between spectral index, $\alpha$, and difference of spectral index between lobes, d$\alpha$, and the fundamental parameters.}
\label{specsr}
\end{table}

\begin{table}
\begin{center}
\begin{tabular}{ccccc}
\hline
 & \multicolumn{4}{c}{Eigenvector}\\
 & 1 & 2 & 3 & 4\\
 \hline
 $z$ & 0.64 & -0.25 & 0.06 & -0.73\\
 $P$ & 0.58 & -0.16 & 0.53 & 0.60\\
 $D$ & -0.47 & -0.12 & 0.82 & -0.31\\
 $\alpha$ & -0.21 & -0.95 & -0.21 & 0.13\\
 \hline
 Variation & 45\% & 25\% & 21\% &10\%\\
 \hline
 $z$ & 0.56 & -0.08 & 0.54 & -0.62\\
 $P$ & 0.55 & 0.44 & 0.26 & 0.66\\
 $D$ & -0.40 & 0.84 & 0.19 & -0.30\\
 d$\alpha$ & 0.48 & 0.30 & -0.77 & -0.29\\
 \hline
 Variation & 51\% & 21\% & 18\% & 10\%\\
 \hline
\end{tabular}
\end{center}
\caption{PCA results for the spectral index and the fundamental parameters.}
\label{specpca}
\end{table}

The dependence of the spectral index. $\alpha$, on the fundamental parameters for radio galaxies of type FRII has been investigated by various authors. \citet{vvw72} and \citet{lo89} find a strong
spectral index--radio luminosity correlation, but no corresponding
correlation with redshift. Interestingly \citet{ak99} find a strong
correlation of spectral index with redshift, but no significant
correlation with radio luminosity. Finally, \citet{brw99} find that $\alpha$ is anti-correlated not only with radio-luminosity,
but also with source size. They found that in general, larger sources
had a steeper spectral index.

In our sample there is no clear trend of the spectral index with any of the fundamental parameters (Tables \ref{specsr} and \ref{specpca}). The difference in the spectral index of the lobes, d$\alpha$, also shows no significant correlations with the fundamental parameters. The absence of the trends claimed in the literature from our sample is probably due to the small sample size. 

\subsection{Rotation measure}
\label{rotation}

\begin{table}
\begin{center}
\begin{tabular}{cccccc}
X, Y & $r_{\rm XY}$ & Student-t & X, Y & $r_{\rm XY}$ & Student-t \\
\hline
$RM$, $z$ & 0.17 & 0.79 & d$RM$, $z$ & 0.50 & 2.65\\
$RM$, $P$ & 0.27 & 1.31 & d$RM$, $P$ & 0.23 & 1.06\\
$RM$, $D$ & -0.05 & -0.23 & d$RM$, $D$ & -0.22 & -1.05\\[1ex]
$\sigma_{\rm RMz}$, $z$ & 0.68 & 4.24 & & & \\
$\sigma_{\rm RMz}$, $P$ & 0.35 & 1.71 & & & \\
$\sigma_{\rm RMz}$, $D$ & -0.30 & -1.46 & & & \\
\hline
\end{tabular}
\end{center}
\caption{Spearman Rank coefficients for the correlations between rotation measure, $RM$,  the difference of the rotation measure between lobes, d$RM$, and the dispersion of the rotation measure, $\sigma _{\rm RMz}$, and the fundamental parameters.}
\label{rmsr}
\end{table}

\begin{table}
\begin{center}
\begin{tabular}{ccccc}
\hline
 & \multicolumn{4}{c}{Eigenvector}\\
 & 1 & 2 & 3 & 4\\
 \hline
 $z$ & 0.64 & 0.24 & -0.05 & -0.73\\
 $P$ & 0.60 & 0.41 & -0.17 & -0.67\\
 $D$ & -0.38 & 0.44 & -0.80 & 0.13\\
 $RM$ & 0.30 & -0.76 & -0.57 & -0.05\\
 \hline
 Variation & 44\% & 24\% & 22\% &11\%\\
 \hline
 $z$ & 0.63 & -0.02 & 0.22 & 0.74\\
 $P$ & 0.58 & -0.12 & 0.49 & -0.64\\
 $D$ & -0.32 & -0.86 & 0.36 & 0.14\\
 d$RM_z$ & 0.40 & -0.49 & -0.76 & -0.13\\
 \hline
 Variation & 46\% & 23\% & 20\% & 11\%\\
 \hline
$z$ & 0.62 & 0.18 & 0.06 & 0.76\\
 $P$ & 0.49 & 0.42 & -0.62 & -0.45\\
 $D$ & -0.32 & 0.89 & 0.33 & 0.03\\
 $\sigma _{\rm RMz}$ & 0.52 & -0.06 & 0.71 & -0.47\\
 \hline
 Variation & 51\% & 22\% & 19\% & 8\%\\
\hline
\end{tabular}
\end{center}
\caption{PCA results for the rotation measure and the fundamental parameters.}
\label{rmpca}
\end{table}

The SR results in Table \ref{rmsr} show that there are no
strong correlations of the fundamental parameters with rotation
measure, $RM$. In every case the correlation is below 95\%
significant. The PCA results in Table \ref{rmpca} also
demonstrate that the $RM$ of a source does not depend on the size,
redshift or the radio luminosity of the source. This is consistent
with our interpretation that the observed $RM$ is caused by the Galactic magnetic field and
not by Faraday screens local to the sources (G04; \citealt{jl87}).

\begin{figure}
\centerline{\includegraphics[width=8.45cm]{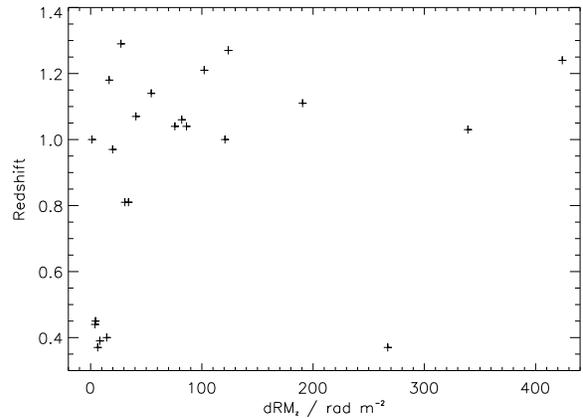}}
\caption{Redshift against difference in rotation measure, d$RM_z$ across the
lobes of a source. d$RM_z$ has been shifted to the sources' frame of
reference.}
\label{zdrm}
\end{figure}

The difference in the rotation measure, d$RM_z$, between the two lobes
of a given source and the dispersion of the rotation measure,
$\sigma_{\rm RMz}$, are probably caused locally to the source and have therefore been corrected to
the sources' frame of reference (G04). There is a d$RM_z$-$z$ correlation
present in the SR results, (see Table \ref{rmsr}), but it
is weaker in the PCA results where the correlation is strongly
reversed in the 3rd eigenvector (see Table \ref{rmpca}). Figure \ref{zdrm} shows that in principal we should find a
strong d$RM_z$-$z$ correlation. However, the presence of the source 3C\,299 at low redshift, but with a large value of d$RM_z$ weakens this trend. The large value of d$RM_z$ for 3C\,299 is due to a lack of polarisation information in one
lobe. Removing this source strengthens the correlation giving
$r_{{\rm d}RM,z} = 0.632$ which corresponds to a significance of 99.98\%.

\begin{figure}
\centerline{\includegraphics[width=8.45cm]{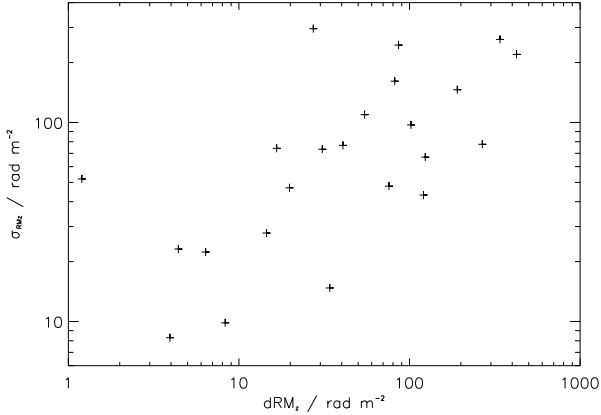}}
\caption{The dispersion of the rotation measure, $\sigma _{\rm RM_z}$, against the difference
in rotation measure across the lobes of a source, d$RM_z$.}
\label{sdrm}
\end{figure}

$\sigma_{\rm RMz}$ measures variations of the
Faraday screen on scales smaller than d$RM_z$, which may in principle be
influenced by the Galactic Faraday screen. However, it is interesting
to note that d$RM_z$ correlates strongly with $\sigma_{\rm RMz}$ (see
Figure \ref{sdrm}). So the two parameters may sample the same, local
Faraday screen even if d$RM_z$ may still contain some Galactic
contribution.

$\sigma_{\rm RMz}$ shows a strong correlation with redshift, $>99.9\%$
significance, and a weak anti-correlation with $D$. The
$\sigma_{\rm RMz}$--$z$ correlation also strengthens with the removal of
3C\,299, becoming $r_{\sigma_{\rm RMz},z} = 0.767$ which corresponds to a
significance exceeding 99.99\%. These correlations are more obvious
when the PCA results are considered (see Table \ref{rmpca}). The first eigenvector contains 51\% of the variation in
the data and shows a strong $z$-$\sigma_{\rm RMz}$ correlation and a weaker
$D$-$\sigma_{\rm RMz}$ anti-correlation. The
$D$-$\sigma_{\rm RMz}$ anti-correlation reverses in the third
eigenvector indicating that it is a considerably weaker trend compared
to the $z$-$\sigma_{RM}$ correlation which only reverses in the last
eigenvector. The PSR results in Table \ref{rmpsr} confirm that
there are no significant correlations with radio luminosity for both
d$RM_z$ and $\sigma_{\rm RMz}$.

\begin{table}
\begin{center}
\begin{tabular}{cccc}
X,Y & Z & $r_{\rm XY,Z}$ & $D_{\rm XY,Z}$\\
\hline
$z$, d$RM_z$ & $P$ & 0.49 & 2.40\\
$z$, $P$ & d$RM_z$ & 0.69 & 3.79\\
$P$, d$RM_z$ & $z$ & -0.19 & 0.86\\[1ex]
$z$, $\sigma _{\rm RMz}$ & $P$ & 0.65 & 3.47\\
$z$, $P$ & $\sigma _{\rm RMz}$ & 0.67 & 3.63\\
$P$, $\sigma _{\rm RMz}$ & $z$ & -0.24 & 1.09\\
\hline
\end{tabular}
\end{center}
\caption{PSR coefficients for correlations of d$RM_z$ and $\sigma _{\rm RMz}$ with the fundamental parameters.}
\label{rmpsr}
\end{table}

In a survey of 27 high redshift sources \citet{prc00} found that
the Faraday rotation was independent of size or radio luminosity, but
also showed that the number of sources with high levels of Faraday
rotation increased with redshift. The results from \citet{prc00} use the $RM$ of a
source in the sources' frame of reference. As noted in G04 and above, the
$RM$ observed in our sources is dominated by contributions from the
Galaxy. Therefore we cannot directly compare our results with those of \citet{prc00}. Also note that the sample of \citet{prc00} contains only sources at considerably higher redshifts than our sample.

The trend reported in \citet{prc00} indicates
that the strength of the magnetic field and/or the density of the
environment is increasing with redshift. Which effect dominates cannot be decided from $RM$ measurements alone. Similarly, we use $\sigma_{\rm RMz}$ and d$RM_z$ which also do not provide us with a direct measurement of the density or the strength of the magnetic field in the vicinity of our sources. However, our indicators depend strongly on the degree of order in the direction of the magnetic field in the local Faraday screen. Thus our results point towards a trend for more chaotic magnetic field structures in the source environments at higher redshift.

\subsection{Depolarisation}

\begin{table}
\begin{center}
\begin{tabular}{cccccc}
X, Y & $r_{\rm XY}$ & Student-t & X, Y & $r_{\rm XY}$ & Student-t \\
\hline
$DM_z$, $z$ & 0.69 & 4.66 & d$DM_z$, $z$ & 0.74 & 5.39\\
$DM_z$, $P$ & 0.69 & 4.65 & d$DM_z$, $P$ & 0.54 & 3.18\\
$DM_z$, $D$ & -0.31 & -1.61 & d$DM_z$, $D$ & -0.31 & -1.60\\
\hline
\end{tabular}
\end{center}
\caption{Spearman Rank coefficients for the correlations between depolarisation, $DM_z$ and  the difference of the depolarisation between lobes, d$DM_z$, and the fundamental parameters.}
\label{dmsr}
\end{table}

\begin{table}
\begin{center}
\begin{tabular}{ccccc}
\hline
 & \multicolumn{4}{c}{Eigenvector}\\
 & 1 & 2 & 3 & 4\\
 \hline
 $z$ & 0.58 & 0.22 & 0.0 & -0.78\\
 $P$ & 0.47 & 0.67 & -0.19 & 0.54\\
 $D$ & -0.44 & 0.61 & 0.64 & -0.16\\
 $DM_z$ & 0.50 & -0.35 & 0.75 & 0.27\\
 \hline
 Variation & 53\% & 22\% & 16\% &10\%\\
 \hline
 $z$ & 0.60 & 0.19 & 0.01 & 0.78\\
 $P$ & 0.48 & 0.68 & -0.22 & -0.52\\
 $D$ & -0.43 & 0.63 & 0.62 & 0.17\\
 d$DM_z$ & 0.49 & -0.33 & 0.75 & -0.30\\
 \hline
 Variation & 52\% & 22\% & 17\% &10\%\\
 \hline
 \end{tabular}
\end{center}
\caption{PCA results for the depolarisation and the fundamental parameters.}
\label{dmpca}
\end{table}

\begin{table}
\begin{center}
\begin{tabular}{cccc}
X,Y & Z & $r_{\rm XY,Z}$ & $D_{\rm XY,Z}$\\
\hline
$z$, $DM_z$ & $P$ & 0.40 & 2.00\\
$z$, $P$ & $DM_z$ & 0.44 & 2.21\\
$P$, $DM_z$ & $z$ & 0.41 & 2.04\\[1ex]
$z$, d$DM_z$ & $P$ & 0.60 & 3.25\\
$z$, $P$ & d$DM_z$ & 0.53 & 2.77\\
$P$, d$DM_z$ & $z$ & 0.05 & 0.23\\
\hline
\end{tabular}
\end{center}
\caption{PSR coefficients for correlations of $DM_z$ and d$DM_z$ with the fundamental parameters.}
\label{dmpsr}
\end{table}

\begin{figure}
\centerline{
\includegraphics[width=8.45cm]{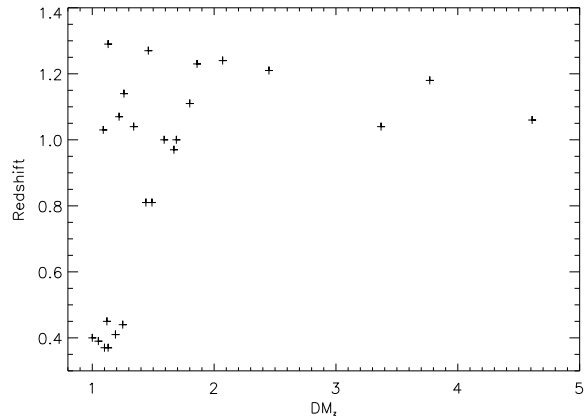}}
\caption
{Redshift against the average depolarisation of a
source. The depolarisation has been shifted to a common
redshift, $z=1$. }
\label{zDM}
\end{figure}

\begin{figure}
\centerline{
\includegraphics[width=8.45cm]{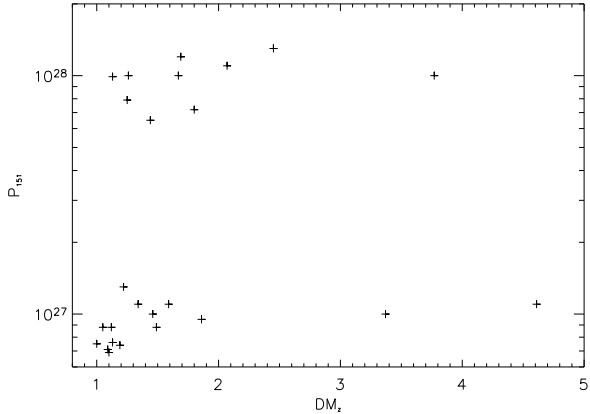}
}
\caption
{Radio-luminosity against the average
depolarisation.  The depolarisation has been shifted to a common
redshift, $z=1$. }
\label{zDM2}
\end{figure}

Depolarisation arises from structure of the magnetic field in the Faraday screen at scales below the telescope resolution. In the previous Section we already found that the measured dispersion of $RM$ is correlated with the redshift of the source, indicating that the small-scale structure in the $RM$ arises from a Faraday screen local to our sources. Therefore we expect that the depolarisation also correlates with redshift in our sample. Clearly we must also correct our measured $DM$ values for redshift effects. In this Section we apply the commonly used model of B66 to adjust $DM$ and d$DM$ to a common redshift of $z=1$ for all our sources to allow for a direct comparison. In Section \ref{tribble} we will investigate alternative models.

There is little
difference between the redshift and the radio luminosity correlations
with the average adjusted depolarisation, taken over the source as a
whole, $DM_z$, (see Table \ref{dmsr}) as both correlations show
significance levels in excess of 99.9\%. There is also an indication
of a $DM_z$-$D$ anti-correlation, but this trend is not significant. We use the Partial Spearman Rank test to determine if
the $P$-$DM_z$ correlation can be explained by the independent
$z$-$DM_z$ and $z$-$P$ correlations (see Table \ref{dmpsr}). Again we find little difference between the redshift
and radio-luminosity correlations. This result is also confirmed by the PCA
results in Table \ref{dmpca}. Figure \ref{zDM} shows that the
higher redshift sources display, on average, a larger degree of
depolarisation and also a much larger spread of depolarisation at any
given redshift compared to their low redshift counterparts. There is little difference in the range of $DM_z$ between sources with very different radio luminosity (see Figure \ref{zDM2}). However, this result may be caused by only two sources of low radio luminosity and exceptionally high values of $DM_z$. Previous studies of the correlation of depolarisation with redshift and/or radio luminosity are also inconlusive. \citet{mt73} found depolarisation to correlate with radio luminosity
whereas \citet{kcg72} found the correlation was predominantly with
redshift. In view of the stronger correlation of $\sigma _{\rm RMz}$ with redshift it is surprising that $DM_z$ does not show a correlation with redshift that is significantly stronger than the correlation with radio luminosity. However, we show in Section \ref{tribble} that this result depends on which model for the local Faraday screen is used to correct for redshift effects.

The difference in the
depolarisation over the source, $dDM_z$, shows a stronger trend with redshift than the overall depolarisation, $DM_z$ (Tables
\ref{dmsr} and \ref{dmpca}). The PSR results, Table \ref{dmpsr}, indicate a significant correlation between redshift and d$DM_z$ at any
given radio luminosity. The corresponding $P$-d$DM_z$ correlation
is considerably weaker at a given redshift.

Table \ref{dmsr} shows a weak $DM_z$-$D$ anti-correlation. The PCA results in Table \ref{dmpca} confirm that this result is not very significant. However, a $DM_z$-$D$ anti-correlation has been reported by \citet[][hereafter P89]{rs73,sj88,prms89}, \citet{bel99} and \citet{isk98}, but was not found by
\citet{jd96}. The observations by \citet{rs73}, \citet{sj88} and
\citet{isk98} are at a lower resolution than our measurements and thus
their different findings may be due to this resolution
difference. The observations by \citet{bel99} are at a higher
resolution, but also at higher frequencies (4.8\,GHz to 8.4\,GHz) and so
we would expect any depolarisation trends to be stronger in this
sample. Both P89 and \citet{jd96} however observe the same frequency range
and with the same spatial resolution as in our sample. Even when our results from
sample B, which are taken at the most similar redshift and
radio-luminosity range to those of P89, are analysed separately, we
still find no trend with size. There is no obvious difference in the selection
of the P89 sample and our sample B, except that P89 chose their sample
from sources which had strong emission lines. By preferentially
selecting sources with strong emission lines P89 have chosen sources with jet axes close to our line of sight. This selection may strengthen any
$D$-$DM$ correlation found. However, we do not have
a convincing physical explanation why our results differ from the ones of P89 because of this selection criterion.

It is usually assumed that radio sources are located in stratified
atmospheres. Therefore a small source will be embedded in denser gas
which acts as a more efficient Faraday screen. As the source expands
the lobes will extend beyond the denser
inner atmosphere, thus reducing the amount of depolarisation observed.
In the beginning of Section \ref{fundamentals} we showed that there is no
significant $z$-$D$ anti-correlation in our sample. Therefore the
lack of a pronounced $DM_z$-$D$ anti-correlation suggests that there
is no significant difference in the source environments at any
redshift as the sources become larger. This may be
evidence that the environments of our sources are relatively
homogeneous with little stratification. The lack of an anti-correlation between the source size, $D$, and d$DM_z$ (see Tables \ref{dmsr} and \ref{dmpca}) is consistent with this idea.

\section{Source asymmetries}
\label{asym}

In Section \ref{fundamentals} we investigated the trends of observables averaged over the entire structure of our sources with respect to redshift, total low-frequency radio luminosity and overall source size. In this Section we study in detail the asymmetries between the lobes of our sources, in particular how differences in the lengths and radio fluxes of lobes of individual sources relate to asymmetries in other observable properties. In the following discussion on possible mechanisms that could give rise to asymmetries between the two lobes of a source, we assume that the two jets of a given source are not significantly different from each other in terms of their energy transport rate.

The angle at which a source is orientated to the line--of--sight affects the projected lobe length. Because of finite light travel times sources orientated at
a small angle will have lobes that appear more asymmetrical than
sources at a large angle, assuming there are no differences between the environments of the two lobes. Assuming a source is oriented at an angle
$\theta$ to the line--of--sight, then the ratio of lobe lengths,
$D_1/D_2$, is given by
\citep[e.g.][]{lr79}
\begin{equation}
\frac{D_1}{D_2}=\frac{1+ \left( v_\circ/c \right) \; {\rm cos} \theta}{1-\left(v_\circ/c \right)\;
{\rm cos} \theta},
\end{equation}
where $v_\circ$ is the velocity at which the hotspot is
moving away from the nucleus and $c$ is the speed of light.  The orientation of a source will also induce asymmetries in the the rotation measure and the depolarisation \citep[e.g.][]{gc91}.  In a simple orientation model
emission from the lobe pointing away from the observer will have a
longer path length through the local Faraday screen and thus any
depolarisation measurements of this lobe would be larger than the lobe
pointing towards the observer, which is referred to as the Laing-Garrington
effect. 

The motion of material in the radio hotspots at the end of the lobes of FRII-type sources is likely to be relativistic. Relativistic beaming can therefore also cause asymmetries in a source, independent of path
length through the Faraday screen. As the
hotspot becomes increasingly more beamed, i.e. $\theta\rightarrow0$,
the lobe ratio increases. The flat radio spectrum of the hotspot will begin to dominate the flux
and spectral index in the beamed lobe. Thus as relativistic beaming becomes more
dominant in a source, the beamed lobe will become brighter, longer and
its spectrum flatter than the receding lobe. \citet{lp91b} also find for sources in their sample that the
least depolarised lobe has a flatter spectrum.

Finally, differences in the environments of the two lobes may also introduce asymmetries. For example, theoretical models for the evolution of radio lobes predict that denser environments lead to shorter and brighter lobes with steeper spectral indices \citep[e.g.][]{kda97a}. A jet with a given energy transport rate will expand more slowly in a dense environment compared to the situation in less dense surroundings. The pressure in the lobe as well as the strength of the magnetic field will be higher. The increased strength of the magnetic field will lead to increased energy losses of the radiating relativistic electrons, resulting in a steeper spectral index. Thus a source situated in an environment with an asymmetrical density distribution will develop asymmetrical lobes. In the following we will use the lobe asymmetries observed in our sample in an attempt to distinguish between these possibilities.

 In Section \ref{fundamentals} we only used the overall physical size of our sources, $D$. We now need to measure the lengths of individual lobes. In about half of our sources we do not detect a radio core at our two observing frequencies. In these cases, wherever possible, we used
previously published maps to establish the core position. However, in several sources, most notably 3C\,16, no core is reported in the literature. We estimated the location of the core
position by using the most likely core position from the literature
\citep{lp91}. This approach is not ideal, but as our
maps are not of very high resolution, the values calculated for the
lobe length will only be approximations to the true lengths in any case. 

In Section \ref{fundamentals} we adjusted the measured depolarisation to a common redshift of $z=1$. For this correction the we had to assume a specific underlying structure of the Faraday
screen. Here we compare the two lobes of a given source and so we can use the uncorrected depolarisation to avoid any model dependence. This choice is consistent with previous studies (P89;\citealt{lp91b,gc91}) which also used the observed, uncorrected
depolarisation. 

\subsection{Lobe length and radio flux}

\begin{figure}
\centerline{\includegraphics[width=8.45cm]{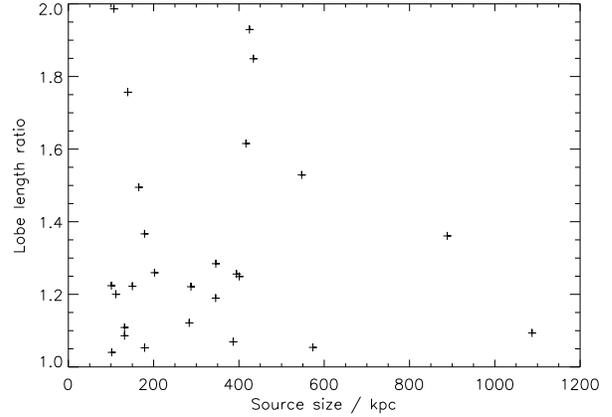}}
\caption
{Plot of the lobe ratio against the physical size of the entire radio structure of our sources.}
\label{lensiz}
\end{figure}

\begin{figure}
\centerline{\includegraphics[width=8.45cm]{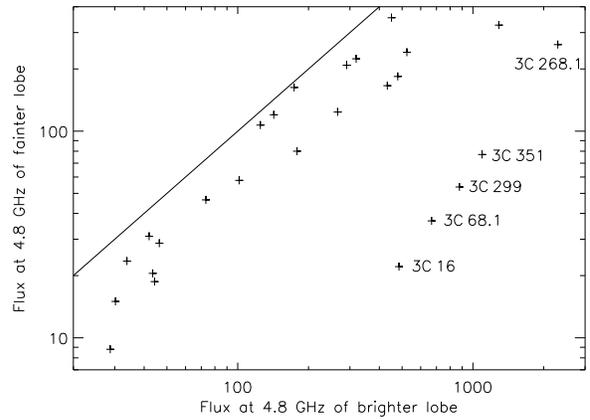}}
\caption
{Plot of the flux of the fainter lobe against the brighter lobe at 4.8\,GHz. The line shows the location of sources for which the luminosities of the two lobes are identical.}
\label{F1F2}
\end{figure}

\begin{figure}
\centerline{\includegraphics[width=8.45cm]{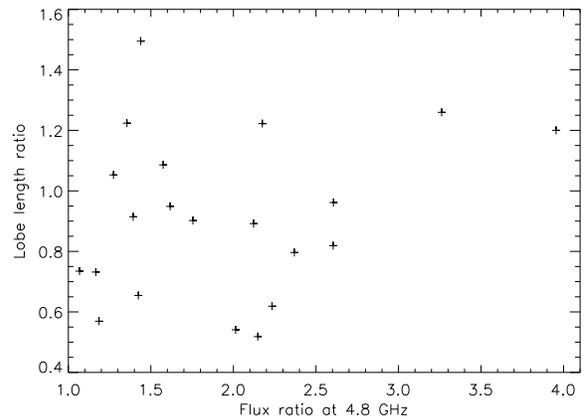}}
\caption
{Plot of the lobe ratio against the flux ratio at 4.8\,GHz, omitting the sources with an exceptional flux asymmetry, 3C\,299, 3C\,268.1, 3C\,16, 3C\,351 and 3C\,68.1.}
\label{FLr}
\end{figure}

Figure \ref{lensiz} shows the ratio of lobe lengths as a function of overall physical size of the sources. Most of our sources are quite symmetric and there is no tendency for smaller sources to show larger lobe length asymmetries than larger sources. 

We use the radio flux at 4.8\,GHz in our study of the source asymmetries, since it has lower noise levels than the 1.4\,GHz data (G04). Figure \ref{F1F2} demonstrates that for all sources the flux of the brighter lobe is tightly correlated with the flux of the fainter lobe. In a few cases the flux of the fainter lobe is
much smaller than the flux of the brighter lobe. Three of these very asymmetrical sources are quasars,
3C\,351, 3C\,68.1 and 3C\,16, the jet axes of which are presumably closely aligned with our line of sight probably leading to the large flux asymmetry. The radio galaxies 3C\,299 and 3C\,268.1 also show a large flux asymmetry. The pronounced asymmetry of 3C\,299 was also noted by \citet{msb95}, who suggest that despite its low redshift the source shows characteristics of a high redshift ($z>$1)
radio galaxy. We also find 3C\,299 to have a $\sigma _{\rm RM}$ comparable to the high redshift samples. 

\citet{hs02} noted that
3C\,299 has lobes of very different lengths. In fact, it is the source with the largest length asymmetry in our sample.  The other four sources showing a large flux asymmetry do not have unusual lobe ratios. However, it should be noted that the cores in 3C\,68.1 and 3C\,16 have not been detected in our observations or previously. \citet{bhl94} do marginally detect a core in
deeper observations of 3C\,68.1 and we have used their core position in determining the lobe lengths of this source. To further aggravate
the problem, both, 3C\,16 and 3C68.1, have very faint lobes and so it is
hard to determine where exactly the lobe terminates. However, it is
worth noting that when the 5 unusual sources are removed from consideration there is no significant trend for the sources with a large flux ratio to also show strong length asymmetries (see Figure \ref{FLr}). 
This argues against the interpretation that the source asymmetries are caused by differences in the gaseous environments of the two lobes of a source. 

We do not find any trends of the flux asymmetries with either redshift or radio luminosity.

\subsection{Spectral index}

\begin{figure}
\centerline{\includegraphics[width=8.45cm]{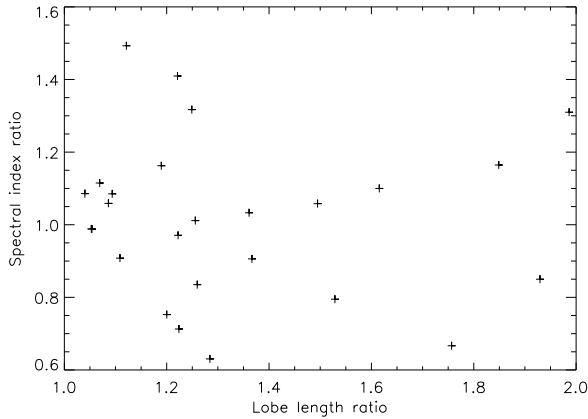}}
\caption
{Plot of the spectral index ratio against the lobe ratio.}
\label{speclen}
\end{figure}

\begin{figure}
\centerline{\includegraphics[width=8.45cm]{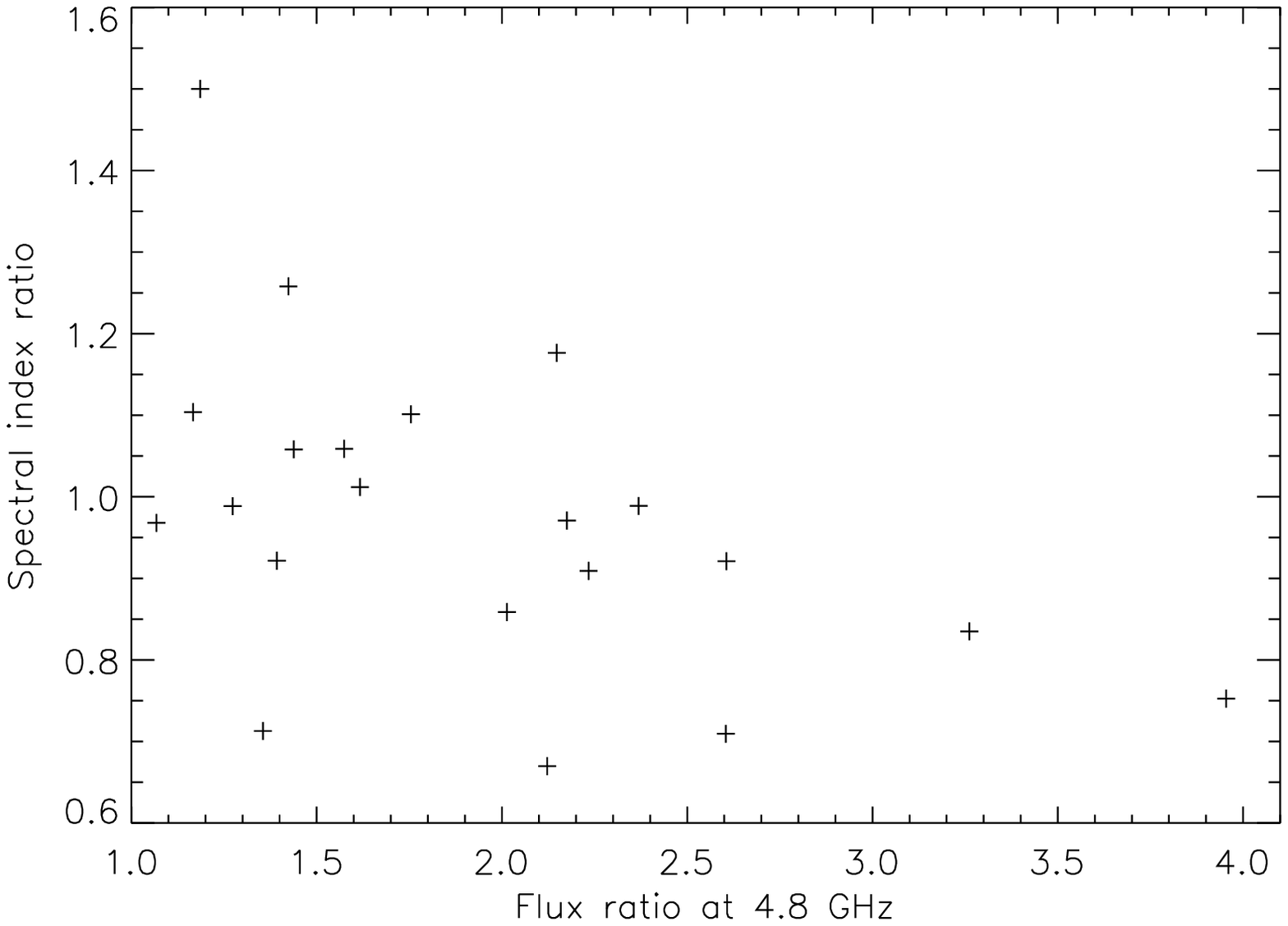}}
\caption
{Plot of the spectral index ratio against the flux ratio at 4.8\,GHz, omitting the sources with an exceptional flux asymmetry, 3C\,299, 3C\,268.1, 3C\,16, 3C\,351 and 3C\,68.1.}
\label{specflu}
\end{figure}

\begin{figure}
\centerline{\includegraphics[width=8.45cm]{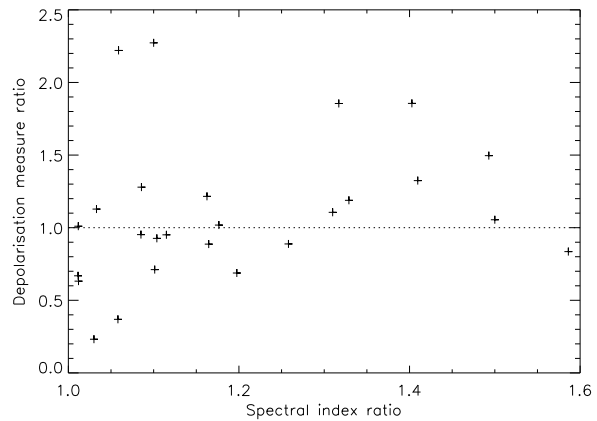}}
\caption
{Plot of the depolarisation measure ratio against the spectral index ratio. Sources above the dotted line follow the trend proposed by \citet{lp91b}.}
\label{specdm}
\end{figure}

The spectral index ratio shows no correlation with the lobe length ratio (see Figure \ref{speclen}). If differences in the gaseous environment of the lobes were the source of the observed asymmetries, then we would expect that the lobe located in the denser environment is shorter and has a steeper radio spectrum \citep[e.g.][]{kda97a}. This result therefore argues against different lobe environments. However, it is consistent with relativistic beaming of the flat-spectrum hotspot emission, if the lobe length asymmetry is due to only small differences in the gas density of their environments. 

Neglecting the five sources with unusually large flux asymmetries, we find that in sources with a large flux asymmetry the brighter lobe has a flatter spectral index than the fainter lobe (see Figure \ref{specflu}). For more symmetric sources the range of the spectral index ratio is larger with a weak tendency for the brighter lobe to have a steeper spectral index than the fainter lobe. Both results are again consistent with asymmetries arising from relativistic beaming of the hotspot emission. Lobes for which beaming is more important will be dominated by their bright hotspots which have a flat radio spectrum. Relativistic beaming is less important for sources oriented close to the plane of the sky. For these objects small differences in the density of the lobe environments may lead to a brighter lobe with a steeper spectrum located in the denser surroundings. 

\citet{lp91b} study a sample of 13 sources, for 12 of which the lobe with the steeper spectrum is also the most depolarised lobe. Figure \ref{specdm} illustrates that only 14 of our 26 sources are consistent with this trend. However, our findings are consistent with those of \citet{ism01} who also find only 58\% of radio galaxies and 59\% of quasars in their much larger sample to agree with the result of \citet{lp91b}. The distribution in Figure \ref{specdm} suggests that sources with a larger spectral index asymmetry are more likely to follow the proposed trend. Again this is consistent with the asymmetries arising from relativistic beaming of the hotspot emission. In this scenario, the lobe with the steeper spectral index points preferentially away from the observer. The radio emission from this lobe has to traverse a longer path through the surrounding medium which acts as a Faraday screen, leading to greater depolarisation. 

\subsection{Depolarisation}
\label{depol}

\begin{figure}
\centerline{\includegraphics[width=8.45cm]{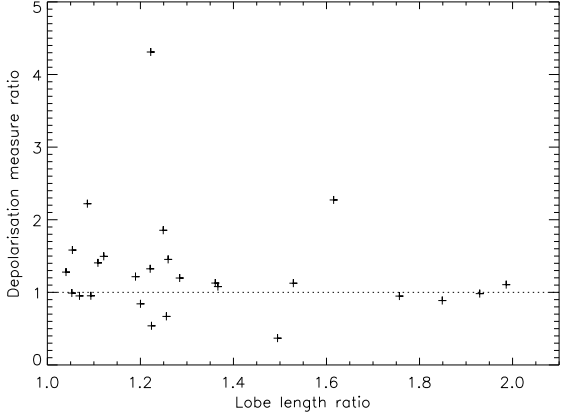}}
\caption
{Plot of the depolarisation measure ratio against the lobe length ratio. For sources above the dotted line the larger lobe is also more depolarised.}
\label{dmlen}
\end{figure}

\begin{figure}
\centerline{\includegraphics[width=8.45cm]{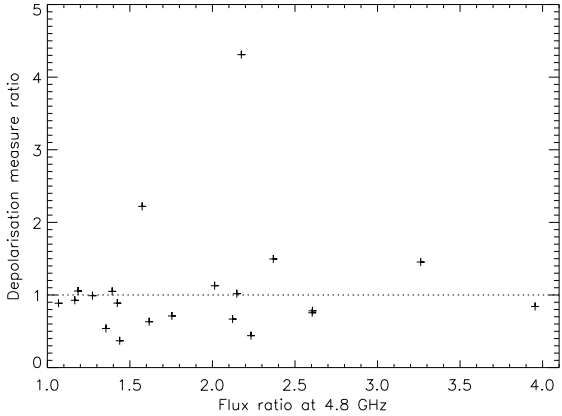}}
\caption
{Plot of the depolarisation measure ratio against the flux ratio at 4.8\,GHz, omitting the sources with an exceptional flux asymmetry, 3C\,299, 3C\,268.1, 3C\,16, 3C\,351 and 3C\,68.1. For sources above the dotted line the brighter lobe is also more depolarised.}
\label{dmflu}
\end{figure}

In Figure \ref{dmlen} we compare the depolarisation measure ratio with the ratio of the lobe lengths. For 16 of our 26 sources the larger lobe is also more depolarised. Our result disagrees with the findings of P89 and \citet{lp91b}, but is consistent with \citet{jd96}. The cause for this difference may be the different sample selection criteria, but the trend is not very strong in any of the studies considered here. 

The same is true for the relation between the depolarisation measure ratio and the flux ratio (see Figure \ref{dmflu}). Neglecting the five sources with large flux ratios, for 13 of the remaining 21 sources the brighter lobe is less depolarised. Although this is consistent with the beaming scenario, the trend is not very significant.

In summary the trends we find in our sample favour relativistic beaming of the hotspots as the underlying mechanism giving rise to the observed source asymmetries. However, none of the trends is very significant and so our results should be treated with caution.

\section{Depolarisation beyond Burn's law}
\label{tribble}

Depolarisation of radio synchrotron emission by a Faraday screen external to the radiation source arises from the inhomogeneity of the magnetic field in the screen combined with limited spatial resolution of the observations. Let the magnetic field and gas density in the Faraday screen be coherent in `cells' of a typical size $s_0$. The physical size of the telescope beam at the position of the source is $t$. For $t\ll s_0$ we resolve the Faraday screen and no depolarisation takes place. In fact, by mapping the rotation measure, we are mapping the structure of the Faraday screen in this case. However, usually $t$ is comparable to or larger than $s_0$ and so one beam contains several cells with different position angles of the polarisation vector, ${\bf p}$. The average polarisation, $< \left| {\bf p} \right| > \equiv p$, is therefore reduced by the partial cancellation of polarisation vectors associated with individual cells within the beam. Clearly, the physical size of the telescope beam depends on the redshift of a source and so $p$ depends in general on $z$ even if the structure of the Faraday screen does not change as a function of redshift. 

In previous sections we have used Burn's law (B66) to correct our measured depolarisation for redshift effects. However, this model of the external Faraday screen only applies if the structure of the Faraday screen is Gaussian and, more importantly, if $s_0 \ll t$, i.e. for cases where the structure of the Faraday screen is completely unresolved. For Cygnus A \citet{cb96} report coherent $RM$ fluctuations on scales of $s_0 \sim 10$\,kpc from observations resolving individual cells. At least in this source we have $s_0 \sim t$ and Burn's law does not apply. 

\subsection{Alternatives to Burn's law}

If we cannot assume that $s_0 \ll t$, we need to take into account the autocorrelation or structure function of the rotation measure cells in the external Faraday screen to derive estimates for the observed $RM$ and polarisation properties of our sources. We follow the approach by \citet[][hereafter T91]{pt91}, assuming a Gaussian random $RM$ field with dispersion $\sigma$ and three different structure functions, $D(s)$, where $s$ measures physical distance projected on the plane of the sky. Also, because of the limit set by the resolution of our observations at wavelength $\lambda$, the $RM$ dispersion we measure, $\sigma _{\rm RMz}$, is not equal to the underlying, `true' dispersion, $\sigma$. In detail we have:
\begin{itemize}
\item Quadratic structure function with
\begin{eqnarray}
D(s) & = & 2 \sigma ^2 \left( s / s_0 \right)^2, \ {\rm for}\ s < s_0 \nonumber\\
D(s) & = & 2 \sigma ^2, \ {\rm for}\, s \ge s_0,
\end{eqnarray}
leading to
\begin{eqnarray}
< \left| {\bf p} \right|^2 > & = & \frac{1-\exp \left(-\beta - \gamma \right)}{1+\gamma / \beta} + \exp \left(-\beta - \gamma \right), \nonumber\\
\sigma^2 _{\rm RMz} & = & \sigma^2 \left\{ 1 - \left[ 1 - \exp \left( - \beta \right) \right] / \beta\right\}.
\label{quad}
\end{eqnarray}
\item Gaussian autocorrelation function with
\begin{equation}
D(s) = 2 \sigma ^2 \left[ 1 - \exp \left( - s^2 / s_0^2 \right) \right],
\end{equation}
leading to
\begin{eqnarray}
< \left| {\bf p} \right|^2 > & = & \beta \exp \left( - \gamma \right) \int_0^1 x^{\beta-1} \exp \left( \gamma x \right) \, {\rm d} x , \nonumber \\
\sigma _{\rm RMz}^2 & = & \frac{\sigma ^2}{1 + 1/\beta}.
\label{gaus}
\end{eqnarray}
\item Power-law autocorrelation function with
\begin{equation}
D(s) = 2 \sigma ^2 \left[ 1 - \left( 1 + 2 s^2 / m s_0^2 \right)^{-m/2} \right],
\end{equation}
leading to
\begin{eqnarray}
< \left| {\bf p} \right|^2 > & = & \frac{m}{2} \beta \exp \left( \beta -\gamma\right) , \nonumber \\
& & \int _1 ^{\infty} \exp \left( - m \beta x / 2 \right) \exp \left( \gamma x^{-m/2} \right) \, {\rm d} x \nonumber \\
\sigma _{\rm RMz}^2 & = & \sigma^2 \left( m \beta / 2 \right)^{m/2} \Gamma \left(1-m/2,m\beta/2 \right) \exp \left( m \beta /2 \right).
\label{power}
\end{eqnarray}
\end{itemize}
Here we used the definitions $\beta = s_0^2 / 2 t^2$, $\gamma = 4 \sigma ^2 \lambda ^4$ and 
\begin{equation}
\Gamma (a,x) = \int_x^{\infty} \exp (-t) t^{a-1}\,{\rm d}t
\end{equation}
for the incomplete $\Gamma$-function. The polarised flux can be obtained from $p^2 = < \left| {\bf p} \right|^2 >$ for a Gaussian telescope beam.

The physical size of the telescope beam at the source location, $t$, is given by the resolution of the telescope and the adopted cosmological model. The quadratic and Gaussian structure functions then have two free parameters, the typical cell size, $s_0$, and the dispersion of the Gaussian random field assumed for the $RM$ structure, $\sigma$. The power-law structure function contains a third parameter, the exponent of the power-law, $m$. T91 suggests $m=2$ in his analysis, but other values may apply in different sources. Since for the power-law model we have $p \propto \lambda^{-4/m}$ (T91), we can estimate $m$ for sources with measured polarised fluxes at widely spaced wavelengths. Fitting this prediction to the observations of a small sample of radio galaxies observed at 1.4, 4.8 and 8.4\,GHz \citep{ag95}, we find $1 \le m \le 4$. In the following we use $m=4$ and $m=2$. 

\subsection{Dispersion of rotation measure}

\begin{table}
\begin{center}
\begin{tabular}{cccc}
 & & $\sigma$ / rad\,m$^{-2}$& $s_0$ / kpc\\
 \hline
 Varying $s_0$ & Quadratic & 530 & $0.15 (1+z)^{4.1}$\\
 & Gaussian & 500 & $0.11 (1+z)^{4.2}$\\
 & Power-law, $m=2$ & 490 & $0.03 (1+z)^{5.1}$\\
 & Power-law, $m=4$ & 460 & $0.08 (1+z)^{4.4}$\\
\hline 
Varying $\sigma$ & Quadratic & $5.0 (1+z)^{4.9}$ & 10. \\
& Gaussian & $5.1 (1+z)^{4.7}$ & 10.\\
& Power-law, $m=2$ & $5.1 (1+z)^{4.5}$ & 10.\\
& Power-law, $m=4$ & $5.1 (1+z)^{4.6}$ & 10.\\
\hline
\end{tabular}
\end{center}
\caption{Summary of best-fitting model parameters for the dispersion of the rotation measure, $\sigma _{\rm RM}$.}
\label{sigfit}
\end{table}

For a given set of model parameters $s_0$ and $\sigma$ we calculate from equations (\ref{quad}), (\ref{gaus}) and (\ref{power}) the expected observed $RM$ dispersion, $\sigma _{\rm RM}$, for each source. We compare this model prediction with the observed value of $\sigma_{\rm RMz}$ and calculate the $\chi^2$-deviation of the model for each source. We then minimise the overall $\chi^2$-value for the entire sample by varying $s_0$ and $\sigma$. In the following we do not aim for a formal best-fitting model. We are only interested in the general form of models that are consistent with the data.

\begin{figure}
\centerline{\includegraphics[width=8.45cm]{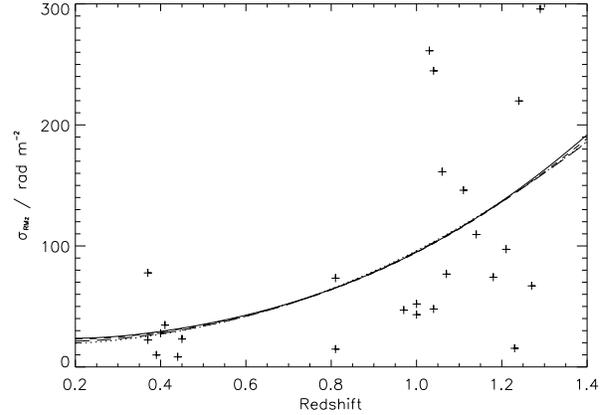}}
\caption{Best fitting models for constant dispersion, $\sigma$, and varying $s_0$. Symbols show the data for our sample. Solid line: Quadratic structure function. Short-dashed line: Gaussian autocorrelation function. Dotted line: Power-law autocorrelation function with $m=2$. Long-dashed line: Power-law autocorrelation function with $m=4$. Model parameters are summarised in Table \ref{sigfit}.}
\label{gausquad}
\end{figure}

Implicitly we assume here that both model parameters do not vary with redshift and are the same for all sources. The expressions for $\sigma _{\rm RMz}$ then lead to $\sigma _{\rm RMz} \sim 0$ for $t < s_0$ and $\sigma _{\rm RMz} \sim \sigma$ for $t>s_0$ with a very steep rise in between. Since $t$ increases as $z$ increases, the very rapid transition of $\sigma _{\rm RM}$ from zero to $\sigma$ must occur at some finite redshift. As Figure \ref{gausquad} shows, our data are not consistent with this behaviour. Incidentally, the usually assumed model of B66 would predict the same behaviour of $\sigma_{\rm RM}$ as the models of T91 with constant $\sigma$ and $s_0$ discussed here. Thus our data are inconsistent with this model. 

Given our data, clearly either $s_0$ or $\sigma$ or both must vary with redshift. Therefore we introduce a third model parameter by setting either $\sigma = {\rm const.}$ and $s_0 = s_{z=0} ( 1+z)^{\mu}$ or, alternatively, $s_0 = {\rm const.}$ and $\sigma = \sigma _{z=0} (1+z)^{\mu}$.  

\begin{figure}
\centerline{\includegraphics[width=8.45cm]{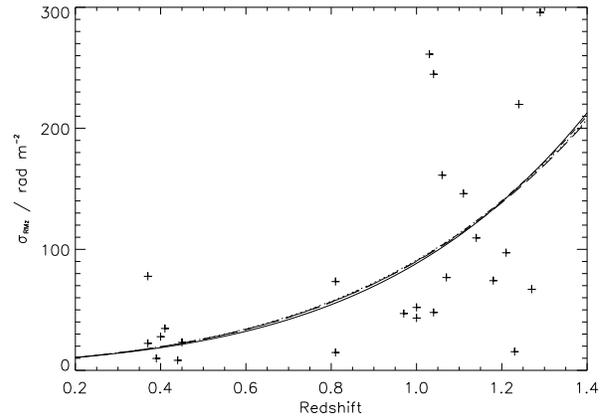}}
\caption{Same as Figure \ref{gausquad}, but for constant $s_0$ and varying dispersion, $\sigma$. Model parameters are summarised in Table \ref{sigfit}.}
\label{powerlaw}
\end{figure}

Figures \ref{gausquad} and \ref{powerlaw} show the model results in comparison with our observations. Note here that our models cannot fit all data points. However, the plots indicate that all of the models are roughly consistent with the data and that either the dispersion, $\sigma$, or the typical cell size, $s_0$, or both must be strong functions of redshift (see Table \ref{sigfit}). None of the models discussed here fits the data significantly better than the others. In fact, the model parameters are quite similar between the models. Therefore measurements of $\sigma_{\rm RMz}$ alone do not rule out any of the models. However, it is also clear that a model varying both parameters $\sigma$ and $s_0$ with redshift will not provide a significantly improved fit. 

\subsection{Depolarisation}

\begin{figure}
\centerline{\includegraphics[width=8.45cm]{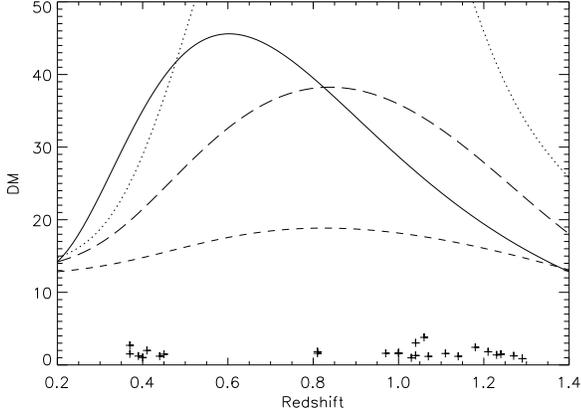}}
\caption{Model predictions for the depolarisation between 1.4\,GHz and 4.8\,GHz compared with our observations. All models assume a constant dispersion $\sigma$ and a varying typical cell size $s_0$. See Table \ref{sigfit} for details of the model parameters. Solid line: Power-law autocorrelation function with $m=2$. Dashed line: Same as solid line, but for $m=4$. Dot-dashed line: Gaussian autocorrelation function. Dotted line: Quadratic structure function.}
\label{s0}
\end{figure}

\begin{figure}
\centerline{\includegraphics[width=8.45cm]{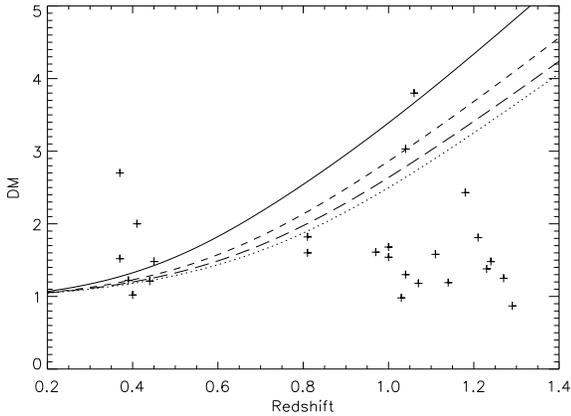}}
\caption{As Figure \ref{s0}, but for constant $s_0$ and varying $\sigma$.}
\label{sigma}
\end{figure}

From equations (\ref{quad}), (\ref{gaus}) and (\ref{power}) we can also derive model predictions for the polarised flux at our two observing frequencies, 1.4\,GHz and 4.8\,GHz. The ratio of the two provides us with a model estimate of the depolarisation, $DM$, which we compare with our observations. We use the best-fitting model parameters of the previous Section, summarised in Table \ref{sigfit}.

Figure \ref{s0} demonstrates that models with a variation of the typical cell size, $s_0$, fail to explain the observations. The expected depolarisation significantly exceeds the observed values in all cases. Although these models provide adequate fits to the observed dispersion of the rotation measure, they do not predict the correct behaviour of the percentage polarisation as a function of observing frequency. Models with a constant cell size $s_0$, but varying dispersion $\sigma$ provide a better description of the observations (Figure \ref{sigma}). The depolarisation predicted by all models at high redshifts is still somewhat higher than that observed. However, models for which the dispersion of the rotation measure is a strong function of redshift are clearly preferred compared to those with a varying typical cell size.

We find that our data are reasonably described by a model assuming a power-law autocorrelation function with exponent $m=4$ and a dispersion varying with redshift as $\sigma = 5.1 \left( z+1 \right)^{4.6}$\,rad\,m$^{-2}$. We now use this model instead of the model proposed by B66 to adjust the measured depolarisation of our sources to the common redshift $z=1$. The adjusted values $DM_z$ show a strong correlation with redshift and a weaker correlation with radio luminosity (see Table \ref{powdmsr}). If the observed depolarisation of our sources is caused by small-scale structure in a Faraday screen local to the sources, then we would expect both $\sigma_{\rm RMz}$ and $DM_z$ to correlate with redshift. In Section \ref{fundamentals} we found only $\sigma _{\rm RM z}$ to show this trend, but not $DM_z$. The results of this Section suggest that the depolarisation model of B66 used in Section \ref{fundamentals} is inadequate in describing our observations. For more accurate models we indeed find that both $\sigma _{\rm RMz}$ and $DM_z$ correlate with redshift. This result strengthens our conclusion in Section \ref{depol} that the predominant effect changing the polarisation properties of our sources as a function of redshift is the increasing disorder in the structure of the magnetic field in the local Faraday screen of the sources. This decrease in the order of the magnetic fields is reflected in the strong increase in the dispersion of the rotation measure fluctuations in the Faraday screens, $\sigma$, with redshift. The change with redshift in the `true' dispersion, $\sigma$, gives rise to the behaviour of the observed dispersion, $\sigma _{\rm RMz}$, and the observed depolarisation, $DM_z$.

\begin{table}
\begin{center}
\begin{tabular}{ccc}
X, Y & $r_{\rm XY}$ & Student-t \\
\hline
$DM_z$, $z$ & 0.68 & 6.1 \\
$DM_z$, $P$ & 0.49 & 3.8 \\
$DM_z$, $D$ & -0.23 & -1.6\\
\hline
\end{tabular}
\end{center}
\caption{Spearman Rank coefficients for the correlations between depolarisation, $DM_z$ and the fundamental parameters. The depolarisation is now adjusted using the best-fitting model with a power-law autocorrelation function with $m=4$. Note that the coefficients are given for depolarisation averaged over the entire source.}
\label{powdmsr}
\end{table}

\section{Summary and conclusions}
\label{conc}

In this paper we investigate in detail radio observations of a sample of 26 extragalactic radio sources of type FRII. The original observations and data reduction are discussed in G04. The observed spectral index, $\alpha$, the structure of the rotation measure, $RM$, and depolarisation, $DM$, of our sources provide us with information on the conditions in the Faraday screens local to the sources. 

We first adopt the cosmological redshift, the low frequency radio luminosity and the physical size of our sources as the `fundamental' parameters. We find that the radio spectral index shows no significant correlations with any of the fundamental parameters. The same holds for the rotation measure. The latter result is consistent with our suggestion in G04 that most of the rotation of the plane of polarisation occurs in our Galaxy. On smaller scales the difference of the rotation measure between the two lobes of each source, d$RM_z$, and the dispersion of the rotation measure, $\sigma _{\rm RMz}$, both correlate with redshift and with each other. Thus we conclude that the variations of $RM$ on small scales is caused by a Faraday screen local to the source. The redshift correlations also imply that the properties of these local Faraday screens change as a function of cosmic epoch. There is no independent correlation of d$RM_z$ or $\sigma _{\rm RMz}$ with radio luminosity or source size.

The small-scale structure of the rotation measure gives rise to the measured depolarisation of the sources. We initially use the model of B66 for the Faraday screens to convert from the measured $DM$ to $DM_z$, which we would measure if all our sources were located at the same redshift $z=1$. The redshift dependence of d$RM_z$ and particularly $\sigma_{\rm RMz}$ should lead to a similar correlation of $DM_z$ with $z$. However, although such a correlation is present in our data, it is not significantly stronger than the independent correlation of $DM_z$ with radio luminosity. We argue in Section \ref{tribble} that the model of B66 is inadequate for describing our data and weakens the correlation between $DM_z$ and redshift. We do not find an anti-correlation of depolarisation measure with size. Some previous studies (\citealt{rs73,sj88}; P89; \citealt{bel99,isk98}) find such an anti-correlation while another study does not \citep{jd96}. In most cases where there is a contradiction between our result and those presented in the literature, the differences can probably be attributed to different telescope resolutions and/or different observing frequencies. However, in the case of P89 and \citep{jd96} we have no ready explanation for the disagreement. 

We also study the asymmetries between the two lobes of our sources and connections between different kinds of asymmetries. The results are somewhat inconclusive, but seem to favour relativistic beaming of the radio emission of the hotspots as the cause for the observed asymmetries. 

We initially employ the commonly used model of B66 for the Faraday screens affecting the polarisation properties of our sources. However, as T91 points out, this `standard' model only applies for observations in which the structure of the magnetic field of the Faraday screen is not resolved. We therefore use the more sophisticated models by T91 to derive estimates for $\sigma _{\rm RMz}$ and $DM_z$ which we then compare with our observations. We find that only models for which the dispersion of the distribution of $RM$ in the Faraday screens is a strongly increasing function of redshift can successfully explain the observations. All our results therefore point to a decrease in the order of the magnetic field in the Faraday screens with increasing redshift. If the structure of the magnetic field reflects the state of the gas in the source environments, then this finding implies that radio-loud AGN at high redshifts are located in more turbulent environments compared with their low redshift counterparts. 

\section*{Acknowledgments}
We would like to thank the referee for very helpful comments. JAG thanks PPARC for support in the form of a postgraduate studentship. CRK thanks PPARC for rolling grant support.

\def\newblock{\hskip .11em plus .33em minus .07em}

\bibliography{crk}
\bibliographystyle{mn2e}

\end{document}